\documentclass[aps,groupaddress,superscriptaddress,reprint,amsmath,amssymb,graphicx]{revtex4-2}
\usepackage{graphicx}
\usepackage{physics}
\usepackage{bm}
\usepackage{bbold}
\usepackage{xcolor}
\usepackage{siunitx}
\usepackage{amsthm}

\usepackage{setspace}
\usepackage[colorlinks,citecolor=blue]{hyperref}
\usepackage[capitalize]{cleveref}

\newcommand{\ie}{i.e.\ }

\newcommand{\der}[2]{\frac{\mathrm{d} #1}{\mathrm{d}#2}}

\begin{document}

\title{
A novel metric for species vulnerability and coexistence in spatially-extended ecosystems
}

\author{Davide Bernardi}
\affiliation{Laboratory of Interdisciplinary Physics, Department of Physics and Astronomy ``G. Galilei", University of Padova, Padova, Italy}
\affiliation{National Biodiversity Future Center, Palermo, Italy}
\author{Giorgio Nicoletti}
\affiliation{International Centre for Theoretical Physics, Trieste, Italy}
\author{Prajwal Padmanabha}
\affiliation{Department of Fundamental Microbiology, University of Lausanne, Switzerland}
\author{Samir Suweis}
\affiliation{Laboratory of Interdisciplinary Physics, Department of Physics and Astronomy ``G. Galilei", University of Padova, Padova, Italy}
\author{Sandro Azaele}
\affiliation{Laboratory of Interdisciplinary Physics, Department of Physics and Astronomy ``G. Galilei", University of Padova, Padova, Italy}
\affiliation{National Biodiversity Future Center, Palermo, Italy}
\author{Andrea Rinaldo}
\affiliation{Department of Civil, Environmental and Architectural Engineering, University of Padova, Padova, Italy}
\author{Amos Maritan}
\affiliation{Laboratory of Interdisciplinary Physics, Department of Physics and Astronomy ``G. Galilei", University of Padova, Padova, Italy}
\affiliation{National Biodiversity Future Center, Palermo, Italy}

\begin{abstract}
\noindent
We develop a theoretical framework to understand the persistence and coexistence of competitive species in a spatially explicit metacommunity model with a heterogeneous dispersal kernel. Our analysis, based on methods from the physics of disordered systems and non-Gaussian dynamical mean field theory, reveals that species coexistence is governed by a single key parameter, which we term competitive balance. From competitive balance, we derive a novel metric to quantitatively assess the vulnerability of a species, showing that abundance alone is not sufficient to determine it. Rather, a species' vulnerability crucially depends on the state of the metacommunity as a whole. We test our theory by analyzing two distinct tropical forest datasets, finding excellent agreement with our theoretical predictions. A key step in our analysis is the introduction of a new quantity - the competitive score - which disentangles the abundance distribution and enables us to circumvent the challenge of estimating both the colonization kernel and the joint abundance distribution. Our findings provide novel and fundamental insights into the ecosystem-level trade-offs underlying macroecological patterns and introduce a robust approach for estimating extinction risks.
\end{abstract}

\maketitle

\section{Introduction}
Biodiverse ecosystems are remarkably complex: countless species interact with one another in environments that vary over space and time, responding to multiple forms of stress. Explaining how these communities assemble and persist has been a constant focus of theoretical ecology, and numerous frameworks have been proposed, each shedding light on different facets of ecosystem organization \cite{Lev2000,sole2006,marquet2014theory,LorDe2013}. Yet, a unifying theory that fully accounts for the large number of species, the spatial structure, and environmental heterogeneity remains elusive \cite{Che2000,may2019stability, choudoir2022framework, custer2022ecological}, underscoring the complexity of the problem as the impact of human activities on biodiversity intensifies \cite{CebEhr2015}. This highlights the pressing need to develop predictive models of biodiversity and ecosystem stability \cite{TilCla2017}, as well as to understand the ecological consequences of heterogeneous spatial structures at a fundamental level \cite{di2020dispersal, diniz2020landscape, chen2020trade, loke2023unveiling,marrec2021bitbol,abbara2023bitbol, denk2025spatial}.

Given our better understanding of models without spatial structures, a common strategy to address the overwhelming complexity of real-world ecosystems is to collapse the spatial degrees of freedom and focusing on effective interactions between species \cite{allesina2012stability,altieri2021properties,gupta2021effective, barongalla2023PhysRevLett,poleygalla2023PhysRevE, hatton2024diversity}. While this approach has yielded macroecological patterns consistent with observations \cite{VolBan2005,Azaele_2016,mallmin2024chaotic,jops2025oDwyer}, it also tends to obscure the impact of key processes like dispersal on localized dynamics \cite{holland2008strong}. Metacommunity theories that explicitly incorporate spatial processes \cite{leibold2004metacommunity} are difficult to compare against observational data of large number of species due to simplifying assumptions of either landscape structures or dispersal processes. Inclusion of spatial heterogeneity through relative non-linearity typically reduces to pairwise comparison rather than being community-wide \cite{chesson2000mechanisms}. These limitations stem from the fact that we lack tools to incorporate non-trivial spatial variation in large and dispersing communities.  As a result, our understanding of how multiple species simultaneously compete, disperse, and persist in realistic environments remains incomplete  \cite{zarnetske2017interplay,logue2011empirical}. Furthermore, several theoretical results for spatial ecosystems are derived with the assumption of fully connected habitat patches, making it hard to assess the effect of spatial heterogeneity \cite{roy2020complex, nauta2024topological, lorenzana2024interactions}.

\begin{figure*}[t]
	\centering
	\includegraphics[width=\textwidth]{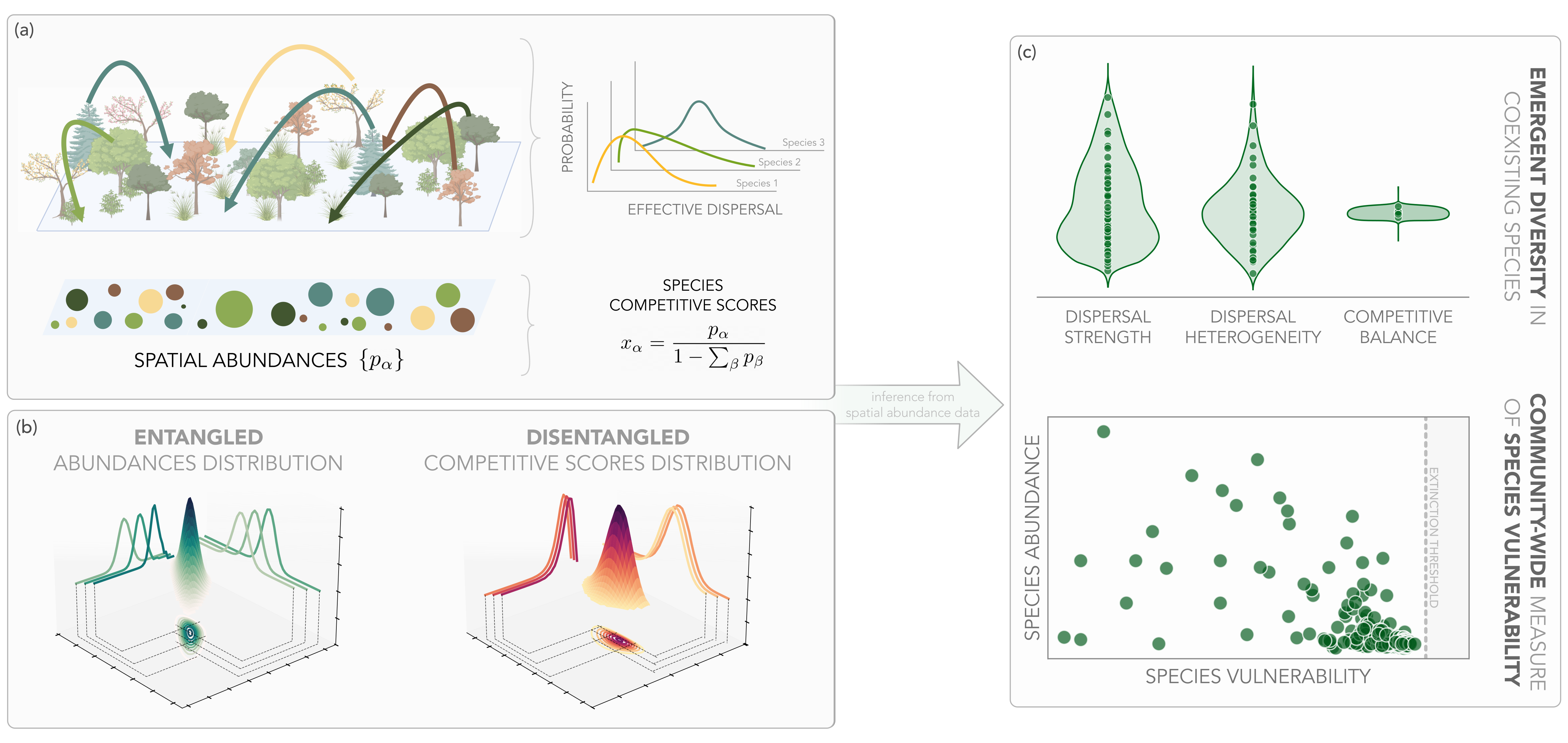}
	\caption{\textbf{A theory of competitive metacommunities with heterogeneous and species-specific dispersal kernels unveils mechanisms underlying coexistence of rare and abundant species and enables assessment of species vulnerability.} a) We consider a model with an arbitrary number $S$ of competitive species and $N$ sites with finite available space / resources. b) The joint distribution of the site occupancy of different species is entangled.  A nonlinear transformation disentangles the abundance joint distribution and yields the species competitive score $x_{\alpha}$. c) Fitting the competitive score distribution enables the extraction of species parameters from tropical forest data. Once single-species parameters are known, the competitive balance and the species vulnerability can be computed from the theory. These two quantities determine whether different competitive species can coexist in a spatially extended, heterogeneous environment, and which species are more vulnerable to disturbances, respectively.}
	\label{fig:Fig1}
\end{figure*}

Here, by utilizing a non-Gaussian Dynamical Mean Field Theory (DMFT) \cite{AzaMar2024}, we develop a comprehensive theoretical framework to understand the conditions for the persistence and coexistence of a spatially embedded metacommunity of competitive species with heterogeneous colonization kernels. We find that the coexistence of different species is determined by a single combination of species parameters, which we term the \emph{competitive balance}. We analytically predict coexistence regions, illustrating how each species’ parameters influence the entire community and deriving a novel metric for species vulnerability. Remarkably, we show that vulnerability is not a sole function of the species' abundance, which depends on the species' dispersal strategy. Crucially, we demonstrate a practical method for testing our predictions using real-world tropical forest datasets. A key insight is that relative abundances are unsuitable for fitting the theoretical predictions. Even in large systems, the joint statistical distribution of relative abundances remains highly entangled, making it extremely difficult to sample adequately from real-world data. Instead, we introduce a new metric, the \emph{competitive score}, defined as the ratio of a species’ relative abundance to the estimated fraction of available space or resources. The joint distribution of competitive scores is expected to factorize—a property derived from a rigorous mathematical argument. Figure \ref{fig:Fig1} shows a conceptual overview of how we combine theoretical tools and observational data to demonstrate coexistence and species vulnerability in tropical forests. Our results have far-reaching implications, revealing links between experimentally accessible abundance patterns and the trade-offs at the ecosystem level that underlie biodiversity. In addition, they offer a novel and more robust approach to estimating species vulnerability. Importantly, our work introduces a new framework for studying ecosystem dynamics—one that balances theoretical tractability with real-world applicability.


\section{Results}
\subsection*{A spatially-explicit metacommunity model for competitive ecosystems}
\noindent We consider a metacommunity model describing the dynamics of $S$ species in a habitat made of $N$ habitat patches, each with a finite amount of colonizable space \cite{PadNic2024}. In each patch, the fraction of space $0 \leq p_{\alpha i} \leq 1$ occupied by species $\alpha$ evolves according to 
\begin{equation}
	\der{p_{\alpha i}(t)}{t} = -e_{\alpha i} p_{\alpha i}(t) + \left(1-\sum_{\beta=1}^{S}p_{\beta i}(t)\right)\sum_{j=1}^{N} K_{\alpha, ij} \, p_{\alpha j}(t),
	\label{eq:model-Grundgleichung}
\end{equation}
where $i=1\dots N$ is the patch index, $\alpha = 1, \dots, S$ is the species index, $K_{\alpha, ij}$ is a species-specific dispersal kernel that describes colonization, and $e_{\alpha i}$ is the local (within patch) extinction rate of species $\alpha$ in patch $i$. The off-diagonal entries of the kernel represent the overall strength of all possible colonization pathways through which a species can move from patch $j$ to patch $i$, \ie the effective rate at which individuals of species $\alpha$ generated in patch $j$ explore the network and eventually colonize patch $i$ \cite{NicPad2023}. Crucially, the term $(1-\sum_{\beta=1}^{S}p_{\beta i})$ quantifies the free space in patch $i$ and acts as a limiting factor for species growth and colonization, introducing competition between species.

Although this model has been shown to describe species survival in fragmented landscapes \cite{HanOva2000, hanski1998, tao2024landscape}, measuring the colonization kernels or the associated dispersal networks in real ecosystems is often challenging \cite{Rogers2019dispersal}. Yet, their importance for ecological dynamics is paramount and well-established \cite{holland2008strong, baguette2013individual, savary2024multiple, savary2024heterogeneous}. To circumvent this issue, here we make the reasonable assumption that, in large-scale ecosystems, entries in the colonization kernel will appear to be randomly distributed. Since $K_{\alpha,ij}$ must be positive, we model the colonization kernel as the Gamma distribution
\begin{equation}
    K_{\alpha, ij} \sim \textrm{Gamma}(\delta_{\alpha}, \beta_{\alpha})
\end{equation}
where $\delta_{\alpha},\beta_{\alpha}$ are the species-specific shape and rate parameters (see Methods). A high $\delta_\alpha$ leads on average to larger entries in the kernel and to a larger spread in the distribution (a stronger dispersal heterogeneity), whereas a high $\beta_\alpha$ tends to reduce colonization between different patches. In particular, both parameters affect the average colonization strength $\ev{K_\alpha} = \frac{\delta_\alpha}{\beta_\alpha}$, whereas the spread of the kernel is controlled by $\delta_\alpha$ through the coefficient of variation $\mathrm{CV}(K_\alpha) = \delta_\alpha^{-1/2}$.

\begin{figure*}[t]
	\centering
	\includegraphics[width=\textwidth]{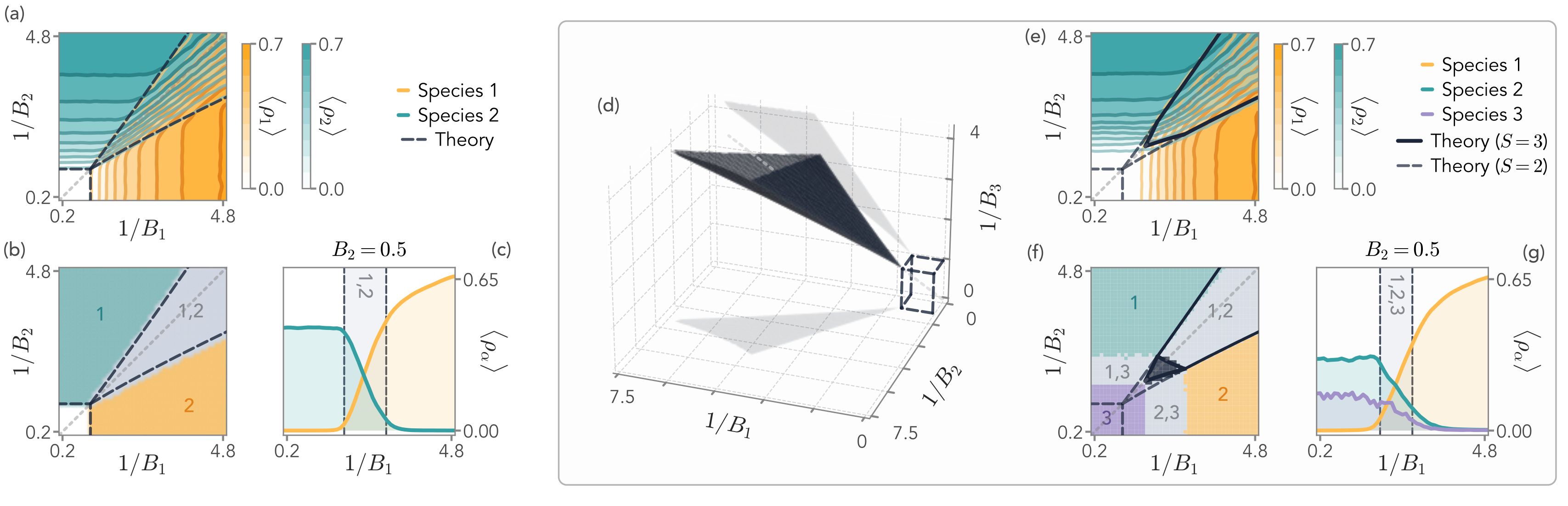}
	\caption{\textbf{Species coexistence is determined by a single key quantity: the competitive balance. Illustrative example with two and three species}. The competitive balance $1/B_{\alpha}$ is a single-species quantity, but its effects can be only determined at a community level, as illustrated by these example phase diagrams for the case of two and three species. a) Abundance of species 1 (orange shading) and species 2 (green shading) as a function of the reciprocal competitive balance; theoretical extinction lines are shown as black dashed lines. b) The phase space is divided into four regions: general extinction (white), only species 1 or 2 survives (green and orange regions, respectively), and coexistence region (gray area). The coexistence region includes but is not symmetrical around the line of equal competitive balance $1/B_1=1/B_2$.  c) Species' abundances as a function of $1/B_1$ for fixed $B_2$, which better illustrates the transitions from survival to extinction and the overlap (gray coexistence region). d) Full phase diagram for three species. Here, the coexistence of all species occurs within the dark volume enclosing the line $B_1=B_2=B_3$ (gray dashed line). e) The presence of species 3, here at the fixed $B_3=0.5$, changes the phase diagram of the first two species: the extinction boundaries for species 1 and 2 are shifted upwards (black solid lines) and the coexistence region for 1 and 2 shrinks.  f) Phase diagram for three species corresponding to a horizontal slice of panel at $B_3=0.5$ d). The coexistence of all three species occurs within the black triangle in the middle. In the other regions, only the indicated species survive g) Slice of panel e), showing the abundance of all three species for fixed values of $B_2$ and $B_3$. Note that changing $B_1$, the competitive balance of species 1, changes the abundances of all species, possibly causing their extinction.}
	\label{fig:Fig2}
\end{figure*}

\subsection*{Competitive balances determine species coexistence and vulnerability}
\noindent In the case of a single species, it is known that the species' persistence depends on whether the local extinction rate is lower than the \emph{metapopulation capacity}, quantified by the largest eigenvalue of the kernel matrix \cite{HanOva2000, ova02, ovaskainen2004metapopulation, NicPad2023}. 
However, when multiple competitive species are considered, each species' metapopulation capacity gives a necessary but not sufficient condition for persistence \cite{PadNic2024}. As detailed in the Methods section, we demonstrate using DMFT that, in the large-$N$ limit, the fate of the metacommunity is governed by
\begin{equation}
    \frac{1}{B_{\alpha}} \equiv\frac{\delta_{\alpha}}{\beta_{\alpha} e_{\alpha}}
\end{equation}
which combines species and habitat-specific parameters. $1 / B_{\alpha} $ represents the \emph{competitive balance} of species $\alpha$.

For a single isolated species, survival is possible if and only if $1/B_{\alpha}>1$, a condition that is equivalent to the classical metapopulation capacity \cite{HanOva2000}. However, when several species are present, the competition for the limited habitat space drastically changes the picture. By exploiting the DMFT solution, we show that, for a large ecosystem, the species that survive and coexist are those that satisfy
\begin{equation}
	W_{\alpha} = \frac{1}{|\Delta|}\biggl(\Delta -  S \frac{\ev{B_{\alpha}} - B_{\alpha} }{\ev{B_{\alpha}}} \biggr) < 0 ,
    \label{eq:species-vulnerability}
\end{equation}
where $\ev{B}$ is the competitive balance averaged across all species, and $\Delta$ is a community-level effective rescaling of the competitive balance with respect to the extinction boundary (see Methods). We name $W_\alpha$ \emph{species vulnerability}: species with a larger vulnerability are at higher risk of extinction. Remarkably, although $W_\alpha$ is defined for the $\alpha$-th species, it depends on the composition of the whole metacommunity through $\ev{B_{\alpha}}$  and $\Delta$, incorporating both its spatial structure and the features of all other species.  Thus, our results explicitly show that the fate of a single species cannot be disentangled from its ecological context. Yet, the species vulnerability $W_\alpha$ is able to encapsulate such complexity in a single measure that, as we will show, can be effectively used to assess diversity and extinction risks in real ecosystems.

\subsection*{Competitive balance in ecosystems with few species}
\noindent To gain insight into the DMFT solution, we now focus on the case where only two species are present ($S = 2$). In \cref{fig:Fig2}a-b we show that the coexistence and extinction transitions (black dashed line) accurately match results from numerical simulations. 
In particular, increasing the competitive balance of species two, $1/B_2$, raises the corresponding value of $1/B_1$ for species one to survive. That is, species one needs to be above a competitive threshold to survive, and as expected the threshold depends on species two. 
The coexistence region always contains the line $1/B_1 = 1/B_2$, which crucially shows that coexistence is possible even in homogeneous environments despite species not having exactly equal values of the competitive balance, contrary to typical competition-colonization tradeoff conditions for coexistence \cite{levins1971regional, levine2002coexistence}. It is interesting to note that the coexistence region is not symmetric with respect to the bisector since the boundaries depend on $B_{\alpha}$ and $\delta_{\alpha}$ separately (see Methods). In particular, \cref{fig:Fig2}c illustrates how changing $1/B_1$ affects both species' abundances at once, shifting the ecosystem between monodominance and coexistence. 

\Cref{fig:Fig2}d shows how the coexistence region changes when a third species is added. The trihedral wedge in the $(1/B_1, 1/B_2, 1/B_3)$ space, inside which coexistence is possible, widens
upon increasing $1/B_{\alpha}$.
With respect to the $S=2$ case, the coexistence boundaries are shifted depending on the competitive balance of species three (\cref{fig:Fig2}e).
Consequently, the region where species 1 and 2 go extinct expands, and they need a higher competitive balance to coexist once the third species is introduced. As shown in \cref{fig:Fig2}f, this effect is due to the fact that the phase space is partitioned by several transition lines, corresponding to all possible combinations of species survival and extinction.
Finally, as we see in \cref{fig:Fig2}g, altering the competitive balance of a single species significantly influences the other species' abundances. Intuitively, introducing the third species narrows the coexistence region of the first two.


\begin{figure*}[th!]
	\centering
	\includegraphics[width = \textwidth]{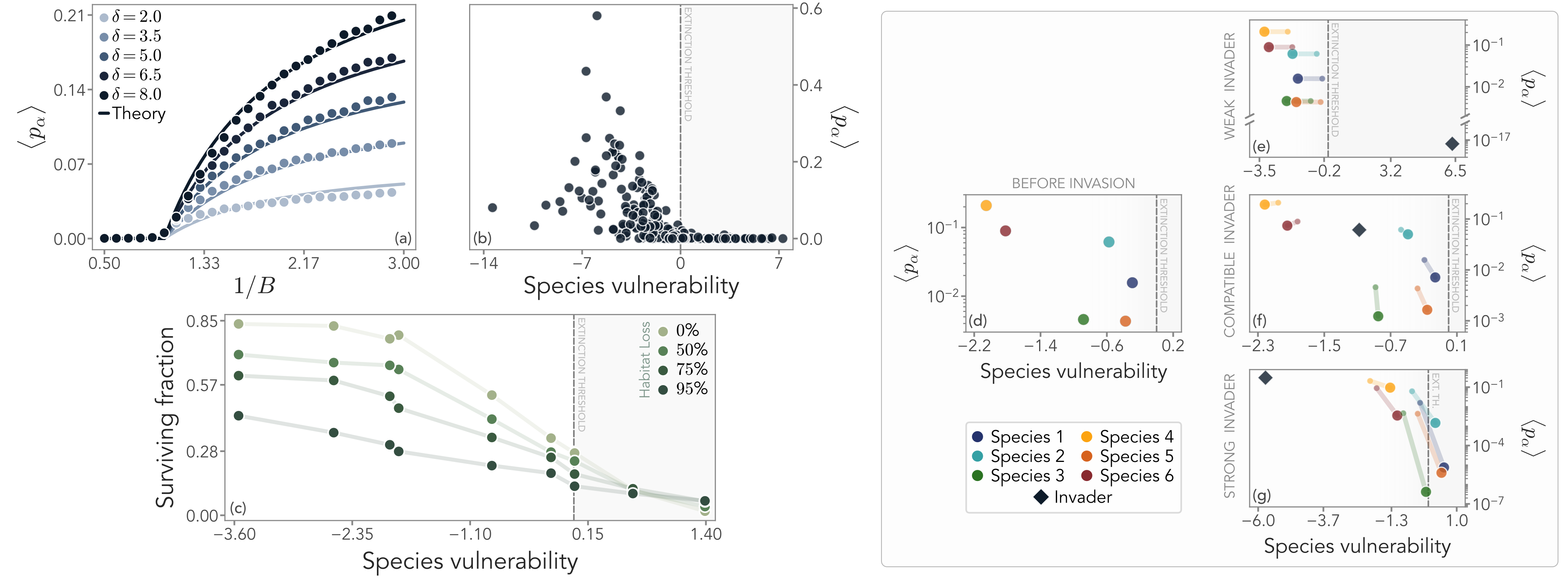}
	\caption{\textbf{Species abundance is neither directly related to the competitive tradeoff, nor to the species vulnerability, which is related to a species' survival chances upon disturbances of various types.} a) The mean species abundance (solid lines: theory; circles: simulations) for a metacommunity of $S=5$ species with identical competitive balance $B_{\alpha} = B$, as a function of $1/B$. Consistent with the theory, the abundances of species with different $\delta_{\alpha}$ are different. b) When the competitive balance is not the same for all species, only species with species vulnerability below zero survive in the long term. The species vulnerability is a community measure, since the competitive balance of all species is needed to compute each species vulnerability. Here we show 50 different metacommunities of $S=6$ species. c) A simplified invasion test on a small metacommunity of six coexisting species, represented in the vulnerability-abundance space. In the following panels, the small circles represent the resident species before invasion, and the larger circles in the same color are the species in the new steady-state after invasion. d) When a species with a much higher vulnerability is added to the metacommunity, the invader goes extinct and the original species are unaffected. e) If the invader's competitive balance is equal to the community's average, the invader can coexist with the original species, which lower their abundance due to the increased competition. f) If the invader has a much lower vulnerability than the existing species, it occupies a large share of the available space causing several extinctions among the more vulnerable species.}
	\label{fig:Fig3}
\end{figure*}

\subsection*{Disentangling survival, coexistence, and species abundances through species vulnerability}
\noindent In order to understand the relation between the species' competitive balances $1/B_\alpha$ and their abundances, we first consider the scenario that all share exactly the same $B_{\alpha} = B$. In this case, all species survive and coexist when $1/B>1$ (see Methods). Yet, this implies neither that all species are equal, nor that they will have the same abundance. In \cref{fig:Fig3}a we show an example ecosystem with five species with the same competitive balance but different parameters $(\delta_\alpha, \beta_\alpha)$. Crucially, we find that their average abundances $\ev{p_\alpha}$ grow differently and become markedly different at large values of $1/B$, as both theory and simulations show. This fined-tuned case suggests that species' survival and coexistence, which depend solely on the competitive balance $1/B$, are not trivially linked to their abundances. Rather, in the Methods we show that in ecosystems with a large number of patches the average abundances become
\begin{equation}
    \ev{p_{\alpha}} = \frac{\beta_{\alpha}}{\bar{\beta}_{\alpha}\left(\vec{\delta}, \vec{\beta}, \vec{e}\right)}
\end{equation}
where $\bar{\beta}_{\alpha}\left(\vec{\delta}, \vec{\beta}, \vec{e}\right)$ has, in general, no simple analytical form in terms of the species parameters and is the solution to a nonlinear implicit integral equation. Therefore, our exact results suggest that species parameters separately determine whether the species survives and its abundance. Furthermore, both depend on the specific composition of the metacommunity - a species that may thrive in one could go extinct in another.


To further investigate this crucial prediction, in \cref{fig:Fig3}b we show the behavior of multiple species’ mean abundances versus their respective vulnerability. We find that most species with a vulnerability above the critical threshold are extinct or near extinction, whereas species with vulnerability far from the threshold persist with non-zero abundances. Remarkably, a strong correlation between each species’ abundance and its vulnerability is only present close to the extinction threshold, but not for low values of vulnerability.
In particular, a perturbative calculation (see Methods) shows that each species' abundance is given by
\begin{equation}
    \ev{p_\alpha} = \frac{|\Delta|}{S \ev{B_\alpha}}\beta_\alpha e_\alpha (-W_\alpha),
    \label{eq:p_alpha_th}
\end{equation}
which is only positive when $W_\alpha$ is negative. Indeed, as manifest in \cref{eq:p_alpha_th}, at a given species vulnerability, the average abundance depends on $\beta_\alpha$ and $e_\alpha$, that is, on the species specific dispersal pattern and local extinction rate. Naturally, the vulnerability $W_\alpha$ deeply constrains the abundance of species with $W_\alpha \approx 0$. However, species with similar vulnerabilities can exhibit vastly different abundances. Conversely, species with similar abundances may have different vulnerabilities, highlighting that abundance alone does not determine a species' risk of extinction. This aligns with ecological thinking, which recognizes that population abundance is not the sole factor in species extinction \cite{gaston2007biodiversity}.

\subsection*{Species vulnerability predicts robustness to habitat loss and ecological invasions}
\noindent Given the spatially-extended nature of our model, we first investigate how the species' vulnerability index provides a measure of a species' susceptibility to habitat disturbances. As a simple example, we model habitat loss by removing a fraction of the patches available to each species, while keeping the model and species parameters fixed. \Cref{fig:Fig3}c presents the fraction of realizations in which each species survives the disturbance, as a function of its vulnerability. As expected, we find that, at all loss levels, the survival fraction decreases as the vulnerability index increases. 

A fundamental property of the vulnerability $W_\alpha$ is that it is not an absolute indicator of extinction risk, but rather a relative measure that intrinsically depends on the parameters of the entire metacommunity. Consequently, adding a new species modifies the vulnerabilities of all resident species. We show a simplified example of a small metacommunity with six species in \Cref{fig:Fig3}c, and focus on three different scenarios of invasion of a seventh species. For a \emph{weak invader}, we add a species with large vulnerability compared to the metacommunity's average (\Cref{fig:Fig3}e). Because of its high vulnerability, this invader goes extinct without altering the abundances of the resident species, though the mere presence of the invader changes the parameters of the metacommunity and thus slightly decreases the vulnerability of each resident. Instead, when the invader has a vulnerability compatible with the metacommunity’s average (\Cref{fig:Fig3}f), it coexists with the original species; all abundances are reduced due to the increased competition, but neither the invader nor any resident species go extinct. Crucially, the vulnerabilities of the new community shift, with some species benefiting from the invasion, and others becoming closer to the extinction threshold $W_\alpha = 0$.  Finally, if the invader has a significantly lower vulnerability (\Cref{fig:Fig3}g), the vulnerability of the resident species increases drastically upon invasion, pushing some of them above the vulnerability threshold and causing extinctions due to the strong competition.

\begin{figure*}[t]
	\centering
	\includegraphics[width=\textwidth]{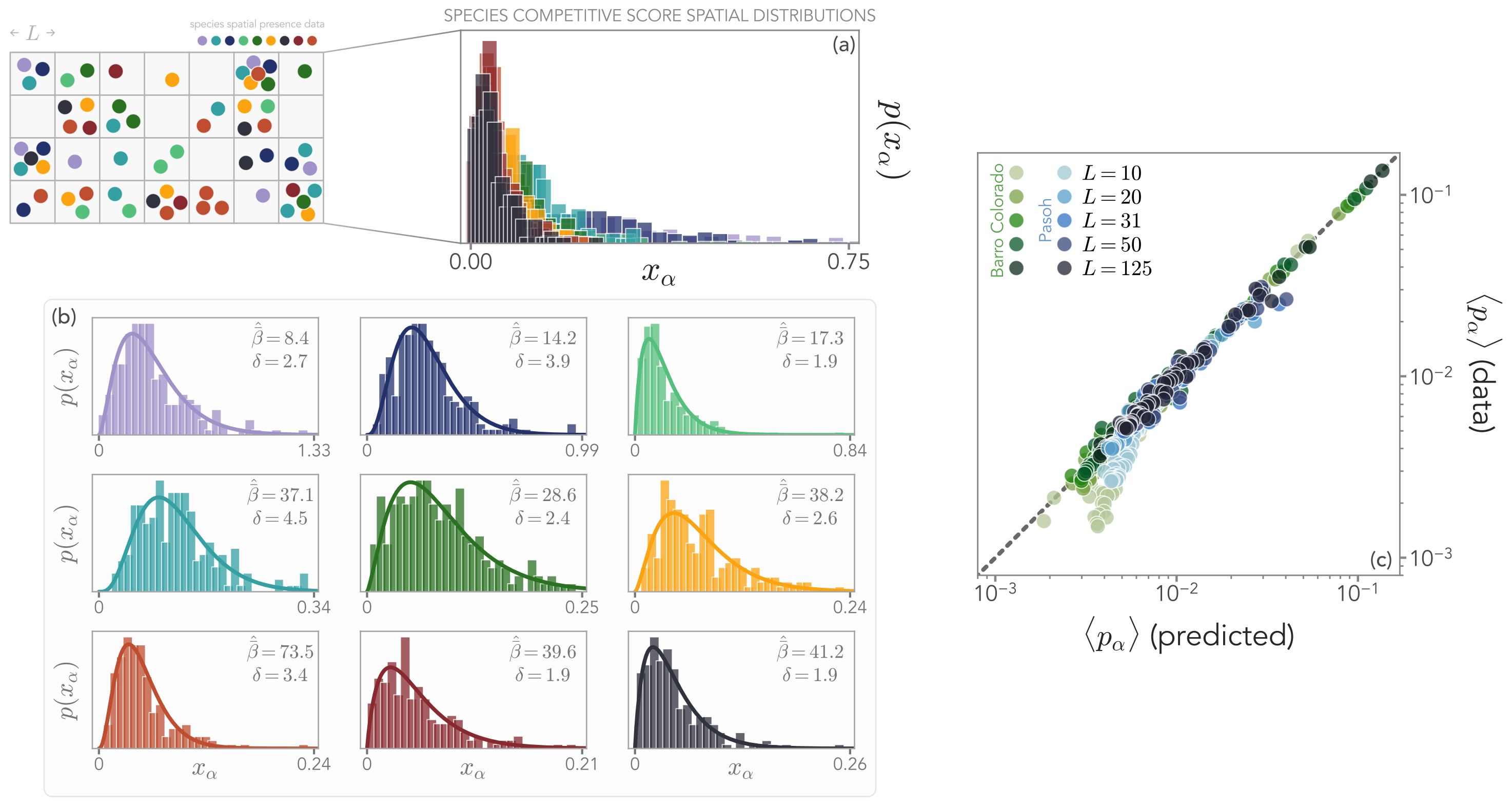}
	\caption{\textbf{Tropical forest data are consistent with the theory.} a) Representative fits of the competitive score of the nine most abundant species of the Barro Colorado Island (BCI) dataset. Fits are performed using the theoretical prediction \cref{eq:x-th-distr}. b) Theoretical prediction for the mean abundance of the 50 most abundant species is in excellent agreement with theoretical prediction for both the BCI (blue circles) and the Pasoh (green circles) tropical forest plots at different level of coarse-graining (shading indicates different coarse graining level, see legend). Only at the smallest coarse graining level ($L=\SI{10}{\meter}$) some minor deviations are seen for low abundances, where sampling noise is maximum.}
	\label{fig:Fig4}
\end{figure*}

\subsection*{Efficient inference from real-world forest data}
\noindent We now test our theoretical predictions on two well-documented tropical forest datasets, the Barro Colorado Island (BCI) \cite{BCI_data} and Pasoh forests \cite{manokaran2004pasoh}. The first steps of our analysis are detailed in \cref{fig:Fig4}a. We divided the 50-ha plots into square subplots of side length $L$, chosen to balance the total number of subplots with the need for sufficient individuals per subplot. To verify that our results did not depend on a specific spatial coarse-graining, we varied $L$ from 10 to 125 \si{\meter}. 
However, fitting the joint abundance distribution of such a high-dimensional dataset is extremely challenging. To avoid this issue, we exploit the fact that the DMFT solution to our model predicts that the joint distribution of the \emph{competitive scores}, defined as
\begin{equation}
	x_{\alpha}\left(\left\{ p\right\}_{\beta} \right) = \frac{p_{\alpha}}{1 - \sum_{\beta} p_{\beta}},
	\label{eq:xstar-main}
\end{equation}
is exactly factorizable in large ecosystems. Thus, the competitive score decouples inter-species dependencies and yields a set of statistically independent variables. As a consequence, we are able to infer the underlying parameters of the high-dimensional joint abundance distribution by fitting the one-dimensional distributions of the competitive scores (see Methods).

The theoretical fits for nine representative species in the BCI dataset are shown in \cref{fig:Fig4}b. 
In all cases, we find an excellent agreement between the data and the distribution fitted from our model. The diversity of shapes in each species’ competitive score distribution highlights their distinct ecological traits, which are well-captured by the theory. Furthermore, beyond predicting the functional form of the competitive score distribution, in \cref{fig:Fig4}b we show that the average abundances predicted by our model using the parameters inferred from the competitive score distributions agree extremely well with data from both forests across all coarse-graining levels. We only start to see deviations from the theoretical expectations for some low-abundance species at finer spatial scales. This is not surprising, given that low abundance and fine spatial resolution are precisely the conditions where sampling noise is most pronounced.

\begin{figure*}[t]
	\centering
	\includegraphics[width=\textwidth]{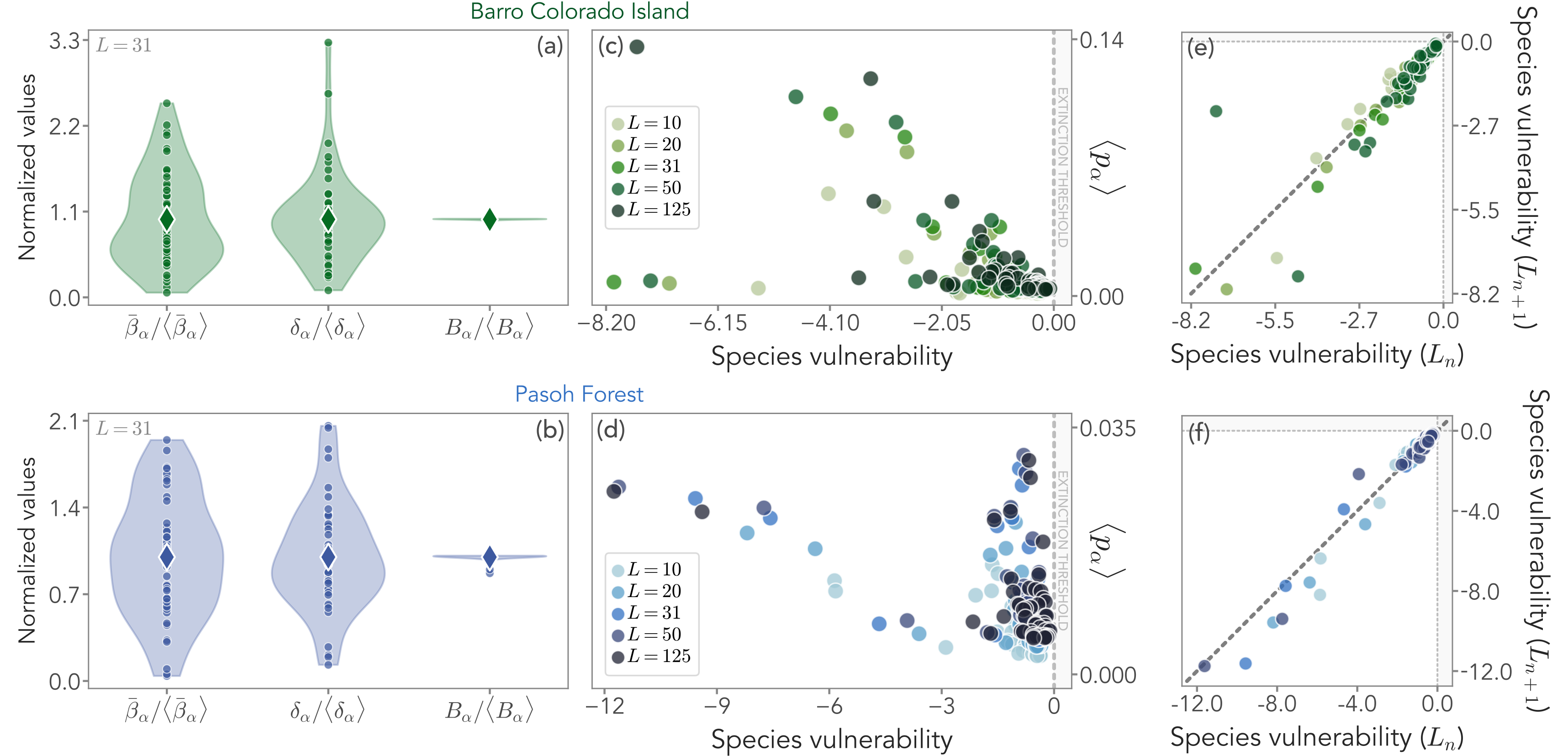}
	\caption{\textbf{Species' parameters are strongly heterogeneous, but their competitive balance is very narrowly distributed, enabling their coexistence. As predicted by theory, species vulnerability is not clearly related to the species abundance.} a) Distribution of species parameters relative to the mean, BCI data. b) Same for Pasoh forest. c,d) Species average abundance vs.\ species vulnerability for all coarse graining levels (see legend) for BCI and Pasoh forests, respectively. e,f) Species vulnerability computed at subsequent coarse graining level pairs, demonstrating consistency of species vulnerability at different levels of coarse graining for BCI and Pasoh datasets, respectively. }
	\label{fig:Fig5}
\end{figure*}

\subsection*{Narrow competitive balances enable coexistence of species with broadly distributed traits}
\noindent Finally, we seek to understand the mechanism by which species coexist in the BCI and Pasoh forests. In \cref{fig:Fig5}a-b, we plot the distribution of the species parameters obtained from the fit of the 50 most abundant species, normalized to the respective mean. We find that both $\delta$ and $\bar{\beta}$, the shape and rate parameters of the species competitive scores (see Methods), are all widely spread, reflecting a high variability of species' effective dispersal strategies. Yet, the picture is markedly different for their competitive balances $1/B_\alpha$. Remarkably, the $B_\alpha$ are very narrowly distributed despite the broad range of individual species parameters. In other words, the competitive balances of the observed species are close to each other, and it is this narrow distribution that enables their widespread coexistence according to our model.

Furthermore, we can compute the species vulnerability according to \cref{eq:species-vulnerability}. The results are once more consistent with the theory - all species vulnerabilities are below the theoretical extinction range, i.e., $W_\alpha < 0$. Crucially, we find once more that abundance and vulnerability are strongly correlated only for species with vulnerability close to zero, close to the extinction threshold. At very low vulnerability, instead, we find a large range $\ev{p_\alpha}$, signaling that species with low average abundance are not necessarily at risk of extinction. These observations are consistent for both forests and all coarse graining levels (\cref{fig:Fig5}c-d). Importantly, we also check that species vulnerabilities remain consistent when computed at different coarse-graining levels - that is, a species that is close to the extinction threshold will be always at risk regardless of the data resolution. In \Cref{fig:Fig5}e,f we report the vulnerability of each species computed at different neighboring levels of coarse-graining, showing that they are strongly correlated and consistent across them. Hence, although the species vulnerabilities cannot be used to compare across different communities as they intrinsically depend on the species at hand, they serve as a within-community and ecologically robust index. 

\section{Discussion}
\noindent In our work, we studied in detail a statistical physics-based model of metacommunity dynamics that accounts for competition and dispersal heterogeneity. Our theoretical predictions, validated through numerical simulations, demonstrate that species coexistence is governed by a complex interplay of colonization, competition, and extinction rates of the entire metacommunity. In the case of a single species, our results reproduce the seminal ones of Hanski and Ovaskainen \cite{HanOva2000} and thus are consistent with standard metapopulation theory. Crucially, we identified a suitably defined competitive balance as a unified basis to characterize large ecosystems. In particular, by employing non-Gaussian dynamical mean-field theory \cite{AzaMar2024} to account for spatial structure, we introduced the species vulnerability $W_\alpha$, that leverages the competitive balances to predict the coexisting metacommunity structure and can be successfully used to assess the extinction hazard of a species. This novel metric summarizes the effect of the entire metacommunity and provides a simple, species-specific criterion for that species' survival in the metacommunity. Crucially, a high vulnerability does not necessarily imply a low abundance, a relation that proves to be true only close to extinction. On the contrary, rare species with low abundances may not necessarily be prone to extinction if their vulnerability in the corresponding metacommunity is low.

We explicitly demonstrated the predictive power of species vulnerability in both theoretical examples, accounting for habitat loss and ecological invasions, and using data from Barro Colorado Island and Pasoh forest. In doing so, we have shown that $W_\alpha$ can be efficiently inferred from large-scale ecological data by exploiting the features of the exact solution of the metacommunity model. Overall, species vulnerability provides a flexible metric to assess both coexistence and extinction risks, accounting for the ecological context. That is, a species with low vulnerability in a given metacommunity may be at risk in another, depending on the species at play and the spatial structure. Our measure of competitive balance effectively acts as a community-wide generalization of the competition-colonization trade-off hypothesis \cite{levins1971regional}. Despite observations of such community-wide tradeoffs \cite{levine2002coexistence,pellowe2014ecology,clark2004fecundity, calcagno2006coexistence, cadotte2007competition}, its implications for coexistence in heterogeneous landscapes and dispersal networks remained unclear \cite{Vellend2017}. In our work, we not only clarify this link, but also demonstrate the implication of such a trade-off in determining vulnerability contingent on the community context.


We stress that our spatially extended theoretical framework allows for an intuitive interpretation of the relationship between abundance and dispersal strategies within the metacommunity. Specifically, at a given vulnerability, species with a larger $\beta_\alpha$ exhibit a higher abundance. This relationship follows from the observation that, at a fixed vulnerability and extinction rate $e_\alpha$, an increase in $\beta_\alpha$ is associated with a corresponding increase in $\delta_\alpha$. Since $\delta_\alpha$ fully determines the coefficient of variation of the dispersal kernel, a larger $\delta_\alpha$ leads to a more uniform dispersal pattern. In other words, species with a high abundance tend to have less spatially variable dispersal, allowing them to colonize a greater number of patches effectively. Conversely, species with more heterogeneous dispersal patterns (broader kernel distributions) will be more subject to crowding effects, at the same level of mean colonization strength. This insight provides a mechanistic link between dispersal heterogeneity and species survival, illustrating how metacommunity structure emerges from individual-level dispersal strategies.

A particularly compelling outcome of our study is the strong agreement between our theoretical predictions and empirical data. This agreement underscores the predictive power of the vulnerability index, while also highlighting its inherent limitations. The vulnerability index does not represent an absolute measure of extinction risk but instead encapsulates a species' fate relative to the entire metacommunity, integrating both dispersal characteristics and environmental heterogeneity. In this sense, it provides a community-level perspective rather than a species-isolated criterion. The consistency of our theoretical predictions across different levels of spatial coarse-graining further reinforces the robustness of the competitive balance as a key determinant of species persistence.

Another  limitation of the species vulnerability index is that it does not provide any information on the timescales over which different species approach their steady state but only predicts the final outcome. In our numerical simulations, we observed that these timescales are highly dependent on species-specific parameters, with some species stabilizing faster than others. This discrepancy suggests that transient dynamics, which are not captured by the vulnerability index, play a crucial role in shaping short- and medium-term ecological interactions. Understanding the factors that drive these variations in transient behavior is an important direction for further investigation. Future work should aim to characterize the dependencies of convergence times on species traits, dispersal properties, and community structure, as this aspects is crucial to gain deeper insights into the mechanisms governing ecological resilience and adaptation in metacommunities, and estimating delayed effects of disturbances, or extinction debts.
    
While our approach focused on the fundamental constraints on coexistence in competitive metacommunities, real ecosystems exhibit additional complexities such as mutualistic interactions, adaptive dispersal strategies, and spatially correlated disturbances. Future research should explore extensions of this framework to include these additional elements. Overall, our study provides tractable approach for studying metacommunity dynamics by bridging ecological theory and statistical physics, and provides a novel perspective on the fundamental principles governing biodiversity patterns and their resilience to perturbations.

\section{Model and Methods}
\subsection*{A spatially-extended community model for competitive ecosystems}
Our model describes the dynamics of $S$ species in an habitat made of $N$ patches, each with a finite amount of available space.  The fraction of site $i=1\dots N$ occupied by species $\alpha$, $p_{\alpha i}$, evolves according to 
\begin{equation}
	\der{p_{\alpha i}(t)}{t} = -e_{\alpha i} p_{\alpha i}(t) + \left(1-\sum_{\beta=1}^{S}p_{\beta i}(t)\right)\sum_{j=1}^{N} K_{\alpha, ij} \, p_{\alpha j}(t),
	\label{eq:model-Grundgleichung-methods}
\end{equation}
where $K_{\alpha, ij}$ is a species-specific dispersal kernel that describes colonization, and $e_{\alpha i}$ is the local (within patch) extinction rate of species $\alpha$ in patch $i$. The kernel represents the overall influence of all possible colonization processes through which patch $j$ can affect patch $i$, \ie the rate at which individuals of species $\alpha$ generated in patch $j$ explore the network and eventually colonize patch $i$. The term $(1-\sum_{\beta=1}^{S}p_{\beta i})$ represents the free space in patch $i$, which introduces competition between species. The dispersal kernel can be derived from an underlying dynamics with explicit ecological reactions mimicking birth, death, exploration, and colonization \cite{NicPad2023,PadNic2024}.

For real ecosystems, the colonization kernel is experimentally inaccessible. It is, however, reasonable to assume that in large-scale heterogeneous natural systems, the colonization kernel will appear randomly distributed. We will now assume that different habitat patches are statistically equivalent. In other words, we assume ``homogeneous inhomogeneity", a reasonable assumption in the absence of explicit macroscopic natural barriers. Then, under reasonable assumptions for how the distribution of the colonization kernel scales with the system size \cite{supplemental_material}, the coupled system of differential equations can be replaced by an effective equation for each species in a representative patch:
\begin{equation}
	\dot{p}_{\alpha}(t) = -e_{\alpha} p_{\alpha}(t) + \left( 1 - \sum_{\beta=1}^{S} p_{\beta}(t) \right) \eta_{\alpha}(t),
	\label{eq:DMFT-main}
\end{equation}
where $\eta(t)$ is a  stochastic process, the properties of which summarize the dynamics of the coupled system, as we derive  exploiting the framework of a non-Gaussian DMFT \cite{AzaMar2024}. The DMFT yields \cite{supplemental_material} a self-consistency relation for the statistical properties of $p_\alpha(t)$, which are entangled with those of the colonization kernel $K$:
\begin{equation}
	\left\langle \prod_{l=1}^{L} \eta_{\alpha}(t_{l}) \right\rangle_{C} =  {N} \left\langle {K}_{\alpha}^{L} \right\rangle_{C} \left\langle \prod_{l=1}^{L} p_{\alpha}(t_{l}) \right\rangle,
	\label{eq:SC-noise-main-def}
\end{equation}
where $\ev{.}_C$ indicates a $L$-th order cumulant and $\ev{.}$ on the right hand side indicates a moment of order $L$. 

\subsection*{Analytical solution yields competitive score and competitive balance}
We focus on the long-term behavior of \cref{eq:DMFT-main}, and ask which species will survive. The stationary state $\left\{p^*_{\alpha}\right\}$ obeys the equation
\begin{equation}
	0 = - e_{\alpha} p_{\alpha}^{*} + \left( 1 - \sum_{\beta=1}^{S} p_{\beta}^{*} \right) \eta_{\alpha}^{*},
	\label{eq:p-star-main}
\end{equation}
where $\left\{\eta^*_{\alpha}\right\}$ is the stationary state of $\eta_\alpha(t)$. From now on, we will drop the asterisks to reduce notation clutter.
We define the  \emph{competitive score} $x_{\alpha} = \eta_{\alpha} / e_{\alpha}$ 
and solve \cref{eq:p-star-main} for it:
\begin{equation}
	x_{\alpha}\left(\left\{ p_{\beta}\right\} \right) = \frac{p_{\alpha}}{1 - \sum_{\beta} p_{\beta}}.
	\label{eq:xstar-methods}
\end{equation}
 An important property of the competitive score derived from the DMFT is that, in the limit of large-system size, the joint distribution factorizes. To obtain explicit analytic predictions, we further assume that the statistical distribution of the kernel can be well approximated by a Gamma distribution
\begin{equation}
	K_{\alpha, ij} \sim \mathcal{Q}_{\alpha N}(K) \equiv \frac{\beta_{\alpha}^{\frac{\delta_{\alpha}}{N}}}{\Gamma(\frac{\delta_{\alpha}}{N})} K^{-1 + \frac{\delta_{\alpha}}{N}} e^{-\beta_{\alpha} K} \Theta(K),
\end{equation}
where parameters are scaled with $N$ to have a consistent large-system limit. With this choice, we find \cite{supplemental_material} that - assuming that the characteristic function of $\mathcal{Q}$ is not too strongly nonlinear nearby the fixed point solution  - the competitive score is itself Gamma-distributed with different parameters:
\begin{equation}
	x_{\alpha, ij} \sim \mathcal{L}_{\alpha}(x) \equiv \frac{\hat{\bar{\beta}}_{\alpha}^{{\delta_{\alpha}}}}{\Gamma({\delta_{\alpha}})} x^{-1 + {\delta_{\alpha}}} e^{-\hat{\bar{\beta}}_{\alpha} x} \Theta(x),
    \label{eq:x-th-distr}
\end{equation}
where $\hat{\bar{\beta}}_{\alpha} = \bar{\beta}_{\alpha} e_\alpha = \beta_{\alpha} e_{\alpha} / \ev{p_{\alpha}}$. To find $\bar{\beta}_{\alpha}$ we start from the following  self-consistency condition for the mean abundances, which originates from \cref{eq:SC-noise-main-def}, evaluated in the stationary state
\begin{equation}
    \ev{p_{\alpha}} = \int \prod\limits_{\mu=1}^{S} \left[ d\eta_{\mu} \, \mathcal{L}^{(\mu)} (\eta_{\mu}) \right] \frac{{\eta_{\alpha}}/{e_{\alpha}}}{1 + \sum_{\gamma} {\eta_{\gamma}}/{e_{\gamma}}}.
    \label{eq:self-consistent-mean}
\end{equation}
Manipulating this last equation  we can show  \cite{supplemental_material} that $\{\bar{\beta}_{\alpha}\}$ are linked to the competitive balance (defined as $1/B_\alpha$) by the following system of integral equations:
\begin{equation}
	B_{\alpha} \equiv \frac{\beta_{\alpha} e_{\alpha}}{\delta_{\alpha}} = \int_{0}^{\infty} d\lambda \, \frac{e^{-\lambda
		- \sum_{\gamma} \delta_{\gamma} \ln \left(1 + {\lambda}/({{\bar{\beta}}_{\gamma} e_{\gamma})}\right)}}{ \left.1 + {\lambda}/({{\bar{\beta}}_{\alpha} e_{\alpha}})\right.}.
	\label{eq:B-alpha-main}
\end{equation}
As reported in the main text, $B_{\alpha}$ is the main quantity that determines whether species $\alpha$ can survive and coexist with other species. A first, necessary condition  for species $\alpha$ to survive is $B_{\alpha} < 1$ or, equivalently, $1/B_{\alpha}>1$. In the presence of other species, however, the critical value at which the species can survive changes, and determining whether a specific set of $B_{\alpha}$ can coexist requires determining a complex set of critical boundaries (see next subsection).

The joint probability distribution of the abundances $\{p_{\alpha}\}$ can be obtained from \cref{eq:x-th-distr} through a change of variables
\begin{equation}
\mathcal{P} (\vec{p}) = \left| \det \left( \frac{\partial \eta_{\gamma}}{\partial p_{\beta}} \right) \right| \prod_{\alpha} \mathcal{L}^{(\alpha)} (\eta_{\alpha}),
\label{eq:P-joint}
\end{equation}
where 
\begin{equation}
\eta_{\alpha} = \frac{p_{\alpha} e_{\alpha}}{1 - \sum_\beta p_{\beta}}.
\label{eq:eta-vs-ps}    
\end{equation}
\Cref{eq:P-joint,eq:eta-vs-ps} make clear that the set of $\{p_{\alpha}\}$ are not independent. Finding an explicit expression for the marginal distributions of $\ev{p_{\alpha}}$ is challenging, since the integration cannot be carried out explicitly and numerical integration becomes challenging when the number of species $S$ is large.

\subsection*{(Co)existence transition boundaries}
As mentioned above, the simple persistence condition $B_{\alpha}<1$ holds only for the species in isolation. When other species are present, the transition boundary depends on the competitive balance of all species. Specifically, the critical values of $\vec{B}$ marking the transition of species $\alpha$ from extinction to survival when all other $S-1$ are present form a $S-1$ dimensional manifold described by the following parametric equations  \cite{supplemental_material}
\begin{align}
B_{\varphi} &= \int_{0}^{\infty} d\lambda \, e^{-\lambda} \prod_{\gamma \neq \alpha}^{S} \left( 1 + t_{\gamma} \lambda \right)^{-\epsilon_{\gamma\varphi}} 
\label{eq:critical-manifolds-1}
\\
\epsilon_{\gamma\varphi} &=
\label{eq:critical-manifolds-2}
\begin{cases}
    \delta_{\varphi} + 1, & \gamma = \varphi \\
    \delta_{\gamma}, & \gamma \neq \varphi
\end{cases}
\end{align}
where $0 < t_{\gamma}< \infty$ are the parametric descriptors of the manifold and $\gamma = 1, 2 \dots \alpha - 1, \alpha +1, \dots, S$. 

We stress that \cref{eq:critical-manifolds-1,eq:critical-manifolds-2} define one of the $S$ transition boundaries for species $\alpha$ \emph{under the assumption that all other species are present}. 
If one or more species are extinct, then they need to be adapted by excluding them from the product term. In the case that exactly one other species is extinct, there are $S$ new transition manifolds (each of dimension $S - 1$). Similarly, if two species are extinct, one obtains a further transition manifolds for each of the $\binom{S}{2} = \frac{S(S-1)}{2}$ possible pairs of extinct species. This pattern continues, leading to a combinatorial explosion of possible transitions to track.

\subsection*{An effective distance from the extinction boundary leads to the species vulnerability}
To circumvent the problem of the combinatorial explosion of transition boundaries, we start from the point in the competitive balance space defined by the condition $B_{\alpha} = B$, \ie the scenario in which the competitive balance of all species are equal to the same value. It can be shown that in this case \cite{supplemental_material} the condition $1 / B > 1$ ensures the coexistence of all species (see \cref{fig:Fig3}a). 
Then, we consider the case of a large ecosystem $S \gg 1$. We make the following ansatz 
\begin{equation}
B_{\alpha} = B_{\alpha 0} \left(1 - \frac{B_{\alpha1}}{S} + \order{\frac{1}{S^2}} \right).
\end{equation}
Then, expanding both sides of \cref{eq:B-alpha-main} as a power series  and matching the leading order contributions leads to the following condition for the survival of all $S$ species \cite{supplemental_material}:
\begin{equation}
B_{\alpha0} = B_{00}; \qquad B_{\alpha1} >  \Delta,
\label{eq:survival-condition}
\end{equation}
where, using $[\cdot]_{\alpha}$ to indicate an average over species
\begin{equation}
\Delta = \frac{ \left[ {\beta}_{\alpha}e_{\alpha} B_{\alpha1} \right]_{\alpha} - B_{00} \left( 1 - B_{00} \right)}{\left[ {\beta}_\alpha e_{\alpha} \right]_{\alpha}}.
\end{equation}
The conditions in \cref{eq:survival-condition} are equivalent to the species vulnerability defined in \cref{eq:species-vulnerability} being smaller than zero.

\subsection*{Forest data analysis}
We divided each 50-ha forest plot into a series of square subplots with side lengths ranging from $L = 10$ \si{\meter} to 125 \si{\meter}. For each subplot, we computed the abundance of all species, $a_{\alpha i}$. To minimize the impact of undersampling, we focused on the 50 most abundant species. We then fitted the parameter $M$, which represents the maximum total number of individuals supported by each subplot, using a greedy deterministic optimization algorithm. This procedure was initialized at the smallest meaningful value,
$M = \max_{i}\Bigl\{\sum_{\beta} a_{\beta i}\Bigr\} + 1$.
Next, we computed the competitive score according to \cref{eq:xstar-methods} and fitted the resulting empirical cumulative distributions using the corresponding theoretical predictions, \cref{eq:x-th-distr}.
To measure the goodness of fit, we calculated a compound negative $\chi^2$ value, which the greedy algorithm aimed to maximize. After 100 consecutive updates with no improvement, the step size in $M$ was increased tenfold. This iterative process continued until no further improvement in the fit was observed.
With this procedure, we obtained the values of $\delta_{\alpha}$ and $\bar{\beta}_{\alpha}$ for each species and coarse graining level, according to \cref{eq:x-th-distr}. Then, we employed \cref{eq:B-alpha-main} to compute each species' competitive balance, from which the value of $\beta_{\alpha}$ was obtained. Finally, we used the relation $\ev{p_{\alpha}} = \beta_{\alpha} / \bar{\beta}_{\alpha}$, which can be derived from \cref{eq:self-consistent-mean}, to test the self-consistency shown in \cref{fig:Fig4}b, and \cref{eq:species-vulnerability} to compute the species vulnerablity shown in \cref{fig:Fig5}b.

\begin{acknowledgments}
We are grateful to the FRIM Pasoh Research Committee (M.N.M. Yusoff, R. Kassim) and the Center for Tropical Research Science (R. Condit, S. Hubbell, R. Foster) for providing the empirical data of the Pasoh and BCI forests, respectively. Original data are available at http://www.ctfs.si.edu.
\end{acknowledgments}

\bibliography{Hanski-DMFT-bib}

\end{document}